\documentclass[10pt,conference]{IEEEtran}
\IEEEoverridecommandlockouts
\usepackage{cite}
\usepackage{amsmath,amssymb,amsfonts}
\usepackage{algorithmic}
\usepackage{graphicx}
\usepackage{textcomp}
\usepackage{xcolor}
\usepackage{listings}
\usepackage[most]{tcolorbox}
\usepackage{ragged2e}
\usepackage{url}
\usepackage{hyperref}
\usepackage{listings}

\lstdefinelanguage{json}{
    basicstyle=\ttfamily\footnotesize,
    numbers=left,
    numberstyle=\tiny,
    stepnumber=1,
    numbersep=5pt,
    showstringspaces=false,
    breaklines=true,
    frame=lines,
    backgroundcolor=\color{white},
    keywordstyle=\color{blue},
    stringstyle=\color{brown},
    morestring=[b]",
    literate=
     *{0}{{{\color{blue}0}}}{1}
      {1}{{{\color{blue}1}}}{1}
      {2}{{{\color{blue}2}}}{1}
      {3}{{{\color{blue}3}}}{1}
      {4}{{{\color{blue}4}}}{1}
      {5}{{{\color{blue}5}}}{1}
      {6}{{{\color{blue}6}}}{1}
      {7}{{{\color{blue}7}}}{1}
      {8}{{{\color{blue}8}}}{1}
      {9}{{{\color{blue}9}}}{1}
}

\def\BibTeX{{\rm B\kern-.05em{\sc i\kern-.025em b}\kern-.08em
    T\kern-.1667em\lower.7ex\hbox{E}\kern-.125emX}}

\begin{document}

\title{An Empirical Evaluation of LLM-Based Approaches for Code Vulnerability Detection: RAG, SFT, and Dual-Agent Systems\\
}

\makeatletter 
\newcommand{\linebreakand}{%
  \end{@IEEEauthorhalign}
  \hfill\mbox{}\par
  \mbox{}\hfill\begin{@IEEEauthorhalign}
}
\makeatother 

\author{
\IEEEauthorblockN{Md Hasan Saju, Maher Muhtadi, Akramul Azim}
\IEEEauthorblockA{\textit{Department of Electrical, Computer, and Software Engineering} \\
\textit{Ontario Tech University} \\
Oshawa, Canada \\
\{mdhasan.saju, maher.muhtadi, akramul.azim\}@ontariotechu.ca}
}

\maketitle

\begin{abstract}
The rapid advancement of Large Language Models (LLMs) presents new opportunities for automated software vulnerability detection, a crucial task in securing modern codebases. This paper presents a comparative study on the effectiveness of LLM-based techniques for detecting software vulnerabilities. The study evaluates three approaches, Retrieval-Augmented Generation (RAG), Supervised Fine-Tuning (SFT), and a Dual-Agent LLM framework, against a baseline LLM model. A curated dataset was compiled from Big-Vul\cite{bigvul} and real-world code repositories from GitHub, focusing on five critical Common Weakness Enumeration (CWE) categories: CWE-119, CWE-399, CWE-264, CWE-20, and CWE-200. Our RAG approach, which integrated external domain knowledge from the internet and the MITRE CWE database, achieved the highest overall accuracy (0.86) and F1 score (0.85), highlighting the value of contextual augmentation. Our SFT approach, implemented using parameter-efficient QLoRA adapters, also demonstrated strong performance. Our Dual-Agent system, an architecture in which a secondary agent audits and refines the output of the first, showed promise in improving reasoning transparency and error mitigation, with reduced resource overhead. These results emphasize that incorporating a domain expertise mechanism significantly strengthens the practical applicability of LLMs in real-world vulnerability detection tasks.
\end{abstract}


\section{Introduction}
 \label{section:introduction}

A software vulnerability is a flaw, caused by weaknesses such as buffer overflows, authentication errors, code injection, or design deficiencies, in the source code that can be exploited by hackers to breach security measures and gain unauthorized access to a system or network \cite{noauthor_cwe_nodate}. This can lead to severe consequences, including data theft, system manipulation, service disruption, and financial loss \cite{noauthor_what_2021}. For example, according to the \emph{Cost of a Data Breach Report 2024} by IBM, the average cost of a data breach is USD 4.88 million which includes the costs of detecting and addressing the breach, disruption and losses, and the damage to the business reputation \cite{noauthor_what_2021}. Vulnerabilities are especially significant in safety-critical systems, where the consequences of exploitation can be catastrophic. As highlighted in \cite{rajesh_vulnerability_2020}, real-time systems like automotive control systems (e.g., anti-lock braking, cruise control) depend on both logical and temporal correctness for faultless operation. A breach in such systems can disrupt timing constraints, leading to missed deadlines and potentially life-threatening failures. Therefore, it is important to detect and mitigate vulnerabilities in a timely manner.

Vulnerability detection involves identifying security weaknesses in software code that attackers could exploit. Conventional detection approaches like rule-based methods and signature-based techniques rely on predefined patterns to spot known vulnerabilities but often fail to detect new or sophisticated threats \cite{guo_comprehensive_2024}. Recent advances in machine learning, especially deep learning, have transformed this field by enabling systems to automatically learn complex patterns from code \cite{guo_comprehensive_2024}. Moreover, the rise of large language models (LLMs) has further enhanced detection capabilities, as these models can analyze code syntax and context to identify vulnerabilities more effectively. LLM models are further customized to improve vulnerability detection through Retrieval-Augmented Generation (RAG) and fine-tuning approaches. However, X.~Du \emph{et al.},\cite{du_vul-rag_2024} and A.~Z.~H. Yang \emph{et al.}\cite{yang_security_2024} highlighted challenges such as false positives and computational costs persist, motivating the exploration of hybrid approaches like Dual-Agent systems.

In this paper, we evaluate and compare the effectiveness of different LLM-based approaches for detecting source code vulnerabilities. The approaches investigated in this paper are RAG, Supervised Fine-Tuning (SFT), and a Dual-Agent system. These approaches are then compared to the performance of the base LLM model. The Dual-Agent system comprises a detector model for identifying vulnerabilities and a validation model for reviewing the first agent's findings. There are some relevant motivations behind this study. Firstly, this study aims to provide a holistic comparison of the three different LLM techniques like RAG, fine-tuning, and Dual-Agent LLMs for vulnerability detection and help researchers or developers decide on the best approach for their needs. Secondly, this paper is the first study to implement and apply a Dual-Agent system in the domain of code vulnerability detection, to the best of our knowledge \cite{li_survey_2024}. Finally, this paper addresses many problems and gaps seen in most vulnerability datasets such as class imbalance, vulnerability coverage limitations and generalizability gaps \cite{guo_comprehensive_2024}, by curating data from multiple sources.

The remainder of this paper is organized as follows: Section \ref{section:related} reviews related work while identifying the gaps that our study addresses; Section \ref{section:methodology} outlines the methodology used to set up the experiment and execute the study; Section \ref{section:results} discusses the results and limitations of the experiment; and Section \ref{section:conclusions} concludes the paper with a summary of the discussion.

\section{Background}
\label{section:related}

Code vulnerability detection has evolved significantly from traditional rule-based methods to modern LLM-based approaches, particularly for safety-critical systems where vulnerabilities can have catastrophic consequences. This section provides context for our research by examining existing LLM-based solutions for vulnerability detection and highlighting how our work addresses dataset gaps in the field.

\subsection{Retrieval-Augmented Generation}
Generally, when an LLM is prompted with a question, it generates a response and reasoning based on the data it was trained on. However, in a RAG \cite{rag} setup, the LLM’s reasoning process is enhanced by providing access to a more up-to-date knowledge base. This means the LLM’s output is informed not only by its inherent reasoning skills but also by the latest information available. When a user submits a query, the LLM first retrieves relevant knowledge from the repository containing the most current data, then refines and ranks that knowledge, synthesizes context, and finally produces the answer. This approach allows the LLM to generate more accurate and timely responses. For example, Vul-RAG \cite{du_vul-rag_2024} uses a knowledge-level framework to detect subtle vulnerabilities, while VulScribeR \cite{daneshvar_exploring_2024} employs RAG for dataset augmentation, improving detection performance.

\subsection{Supervised Fine-Tuning:} 
LLMs are trained on vast amounts of data, which enables them to excel at reasoning and performing general tasks. However, we often require the LLM to handle specialized work. As such, training the model on a specialized dataset can help improve its performance in a specific domain.

\textbf{Parameter Efficient Fine-Tuning: }
There are two ways to provide new knowledge to an LLM through training. One approach is to fully re-train the model, which requires substantial resources and is prohibitively expensive. It also runs the risk of catastrophic forgetting, where the LLM loses previously acquired knowledge and reasoning. A more efficient option is parameter-efficient fine-tuning \cite{peft}, where instead of altering the weights of the pretrained model, we introduce a small set of additional parameters and fine-tune them for our specific use case. This method greatly reduces computational overhead while preserving the original reasoning and knowledge of the model.

\textbf{QLoRA:}
LoRA (Low-Rank Adaptation) \cite{LoRA} addresses the challenge of fully training an LLM by introducing adapters. Instead of training all the model’s weights, it trains only the adapter weights to adapt the model to the target domain using a relevant dataset. This approach provides better reasoning and performance with limited resources. QLoRA is an extension of LoRA that further optimizes memory usage through quantization, storing model parameters as 4-bit or 8-bit values instead of 16-bit or 32-bit. In doing so, QLoRA significantly reduces the computational resources needed to store and fine-tune the model.

Studies like \cite{shestov_finetuning_2025} demonstrate that fine-tuned models outperform smaller architectures, and MSIVD \cite{yang_security_2024} combines fine-tuning with graph neural networks for richer contextual analysis. While RAG excels in leveraging external knowledge, fine-tuning offers task-specific precision.

\subsection{Multi-Agent Systems}
LLM-based multi-agent systems (MAS) are intelligent systems composed of multiple LLM-based agents that collaborate, communicate, and coordinate to accomplish complex tasks \cite{li_survey_2024}. The purpose of these systems is to simulate collective intelligence and mimic human-like social behavior through interaction, reasoning, planning, and adaptation. As \cite{li_survey_2024} outlines, MAS frameworks typically involve a unified framework comprising five components: Profile, Perception, Self-Action, Mutual Interaction, and Evolution. Our Dual-Agent LLM system aligns with this framework by dividing responsibilities between detection and validation agents:
\begin{itemize}
    \item The \textit{Detector Agent} performs initial vulnerability analysis (Self-Action)
    \item The \textit{Validator Agent} provides critical review (Mutual Interaction)
\end{itemize}

While existing MAS research has focused on domains like gaming \cite{li_survey_2024}, our work represents one of the first applications to vulnerability detection, to the best of our knowledge.

\subsection{Dataset Gaps in Vulnerability Detection research}
Despite progress in AI-based vulnerability detection, existing datasets suffer from significant limitations, including severe class imbalance, biased distributions, mislabeling, and poor representation of languages beyond C/C++ \cite{guo_comprehensive_2024}. Our study addresses several of these challenges by curating a balanced and diverse dataset sourced from BigVul and real-world proprietary enterprise codebases (from one of our research partners).  To mitigate the issue of class imbalance, we ensured an equal number of vulnerable and non-vulnerable examples for each of the five selected CWE categories. This step is critical: If, for example, 90\% of the data is non-vulnerable, a model that naively predicts every sample as non-vulnerable would still achieve 90\% accuracy while offering no meaningful detection. By enforcing class balance, we ensure that evaluation metrics reflect true model capability. Additionally, each data instance was reviewed by experts and annotated using structured metadata (e.g., CWE IDs) to ensure high-quality labels. These efforts reduce noise and enhance the generalizability and realism of the dataset, making it more robust and representative for LLM-based vulnerability detection research.


\section{Methodology} \label{section:methodology}
In our work, we explored and compared the capabilities of RAG, SFT, and Dual-Agent LLM in vulnerability detection. We also used a base LLM model as a reference point to assess the performance of the other approaches. In this section, we first describe the dataset used, followed by a detailed explanation of our RAG pipeline, the SFT process, and the Dual-Agent LLM system.

\subsection{Dataset Creation}
To begin with, we created a comprehensive dataset to fine-tune and evaluate the performance of code vulnerability detection using the three LLM approaches and a base LLM model. We collected the data from BigVul and scraped some proprietary repositories from our research partner, GlassHouse, to create our balanced dataset. For each instance, we collected the following features: `CWE ID', `codeLink', `commit\_id', `commit\_message', `func\_after', `func\_before', `lang', `project' and `vul'. For fine-tuning, we used 5000 data instances, and for testing, we used 1000 data instances across five CWE categories. For each CWE, the number of vulnerable and non-vulnerable samples was balanced. As for the vulnerabilities, we considered the following types for our experiment:

\textbf{CWE-119:}
Involves improper restriction of operations within memory buffers, often leading to buffer overflow vulnerabilities.

\textbf{CWE-20:}
 Focuses on improper input validation, which can allow attackers to supply unexpected inputs that compromise security.

\textbf{CWE-399:}
 Pertains to resource management errors, such as failure to properly handle memory or file resources.

\textbf{CWE-264:}
 Covers issues with permissions, privileges, and access controls, potentially allowing unauthorized operations.


\textbf{CWE-200:}
Deals with information exposure, where sensitive data may be leaked to unauthorized parties.

\begin{figure}[t]
\centering
\begin{tcolorbox}[colback=white, colframe=black, title=LLM Classification Prompt, fonttitle=\bfseries, coltitle=white, boxsep=1mm, left=1mm, right=1mm, fontupper=\scriptsize]
\RaggedRight
\scriptsize

You are an expert in cybersecurity and software vulnerability detection. Given a code snippet with modifications, analyze whether the change fixes or mitigates a vulnerability. Use the commit message, function before and after changes, and CVE information (if available) to determine vulnerability.

\begin{enumerate}
    \item \textbf{Analyze the Commit Message} \\
    Does the commit message mention security or vulnerability related terms like 
    \textit{buffer overflow}, \textit{fix}, \textit{CVE}, or \textit{patch}?

    \item \textbf{Compare Function Before and After Changes} \\
    What specific modifications were made to the code?
    Do the changes address known security issues, or do they introduce new risks?

    \item \textbf{Check for CVE/CWE References} \\
    If a CVE is mentioned, what vulnerability does it relate to?
    If a CWE is provided, does the change align with known vulnerability patterns?

    \item \textbf{Assess the Overall Security Impact} \\
    Does the modification fix an existing issue, introduce a new vulnerability, 
    or is it neutral?
\end{enumerate}

Here is the data:

- \textbf{Commit Message}: \{\texttt{sample['commit\_message']}\} \\
- \textbf{Code Before Change}: \{\texttt{sample['func\_before']}\} \\
- \textbf{Code After Change}: \{\texttt{sample['func\_after']}\} \\
- \textbf{CVE ID (if available)}: \{\texttt{sample.get('CVE ID', 'N/A')}\} \\
- \textbf{CWE ID (if available)}: \{\texttt{sample.get('CWE ID', 'N/A')}\} \\
- \textbf{Project}: \{\texttt{sample['project']}\} \\
- \textbf{Programming Language}: \{\texttt{sample['lang']}\}

\textbf{Question:} \\
Does this code change fix or mitigate a vulnerability? 
Answer with \texttt{Vulnerable} or \texttt{Not Vulnerable} and provide reasoning.

\end{tcolorbox}
\caption{Instructional prompt for the LLM Classification task.}
\label{fig:classification_prompt}
\end{figure}

\subsection{Prompt Engineering}
We experimented with various prompts and ultimately arrived at the final version shown in Figure~\ref{fig:classification_prompt}. This version was refined through trial and error. The same prompt was used across all techniques and models to ensure consistency. A separate prompt was used exclusively for the validation agent in the Dual-Agent LLM approach. All prompts are available at the following (see Footnote~\ref{footnotee}).


\subsection{RAG Approach}

\begin{figure}[t]
\centerline{\includegraphics[width=0.51\textwidth]{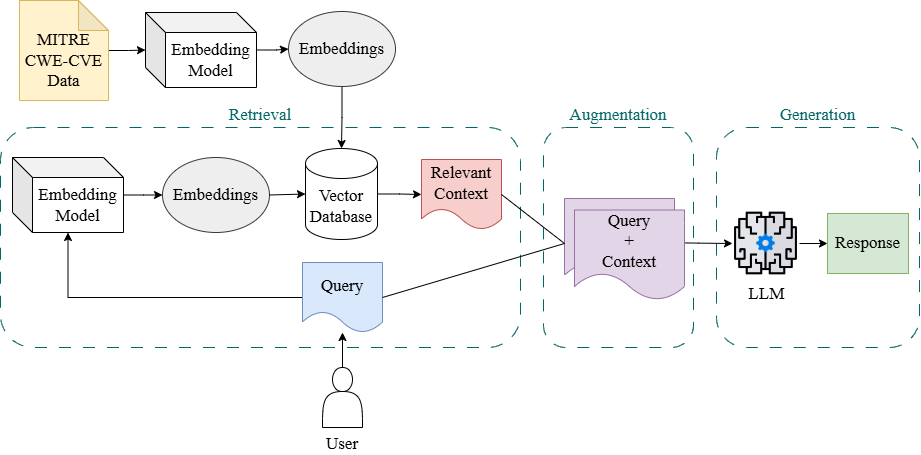}}
\caption{Retrieval-Augmentation Generation Process}
\label{fig:rag}
\end{figure}

The following is a breakdown of the RAG pipeline we set up for our experiment.

\textbf{Data Ingestion:}
 First, we selected the knowledge base for Data Ingestion. We collected all vulnerability-related information from the MITRE Corporation by compiling the knowledge from the Common Weakness Enumeration (CWE) database. The CWE is a standardized list of software and hardware security weaknesses maintained by the MITRE Corporation with support from the National Cyber Security Division of the U.S. Department of Homeland Security \cite{noauthor_cwe_nodate}. The PDF documents collected and compiled from the database were first opened and read page by page using PyMuPDF (fitz). Each page’s text was then passed through a custom regex filter to remove unwanted characters and normalize whitespace, resulting in a cleaner textual representation. This text was combined and subsequently split into more manageable segments (or “nodes”) by a SentenceSplitter, ensuring coherent data chunks. Finally, the pipeline stored the processed data in an internal data structure keyed by each file’s name, allowing for efficient retrieval, resetting, and management of ingested content.
 
\textbf{Data Processing:}
The raw documents were segmented into smaller chunks to facilitate efficient processing. We used a chunk\_size of 512 tokens and a chunk\_overlap of 32 tokens. Smaller chunks enable faster and more accurate retrieval, while overlapping ensures contextual continuity across segments. Each chunk was then passed through an embedding model, with caching and batch sizes carefully managed to optimize retrieval performance. These configurations were stored and reused to ensure flexible and consistent data preprocessing across the system. Below, we added an example of data tokenization: 

\begin{verbatim}
["bool", "IsTextTooLongAt", "(",
 "const", "Position", "&", "position",
 "const", "Element", "*", "element", "=",
 "EnclosingTextControl",
 "…………",
 "return", "toHTMLTextAreaElement",
 "(", "element", ")", "->", "TooLong",
 "return", "false", ";", "}"]
\end{verbatim}

\textbf{Embeddings:}
After the data was tokenized, the knowledge chunks were converted into vector embeddings. These embeddings retain the relevant information from each data chunk. We can then perform a similarity search using these vector embeddings to retrieve our preferred or target context. Finally, we stored the vector data in a high-dimensional vector store called Chroma. Each chunk was embedded to produce a numerical representation in a high-dimensional vector space, capturing semantic relationships within the text. This transformation enables tasks such as similarity searches to locate semantically related chunks quickly. The embeddings themselves were stored in Chroma, along with relevant metadata, making it easy to update, retrieve, and scale the data. Here is an example of knowledge embeddings: 

\begin{verbatim}
[ 0.2173, -0.1359,  0.0661,  …………, 
 -0.0225,  0.3464,  0.0887, -0.1032 ]
... ... ... ..... ....... ........
[ 0.1729,  0.3165, -0.1151,  …………,  
  0.2556, -0.0899,  0.0347,  0.4872 ]
\end{verbatim}

\textbf{Retrieve Related Context:}
 In this stage, when a user submits a query, the query is processed using the same methodology as the knowledge base. The formatted query is then used to search the vector database for relevant data chunks. Our RAG pipeline retrieves the 20 most relevant chunks based on similarity scores, which are subsequently re-ranked and refined. Finally, the context and the query are sent to the LLM to generate a response. This approach enables the LLM to make more informed decisions based on the latest domain data, while also reducing hallucinations. Figure \ref{fig:rag} gives an overall idea of this RAG pipeline.


\subsection{SFT Approach}

\begin
{figure}[t]
\centerline{\includegraphics[width=0.45\textwidth]{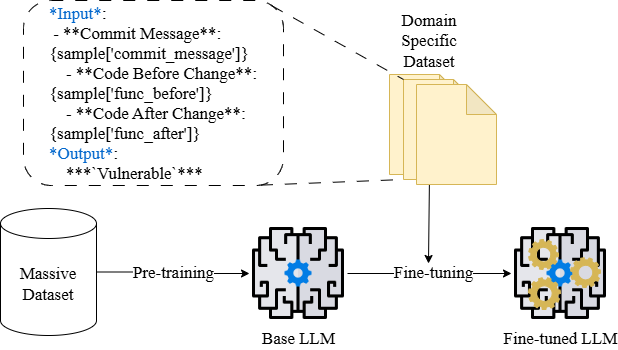}}
\caption{Supervised Fine-Tuning Process}
\label{fig:finetuning}
\end{figure}

Our SFT was implemented with parameter-efficient QLoRA adapters. We trained our model on our vulnerability dataset using a three-part format (instruction, input, and output) for the training data. The instruction portion provided the system directive for the model, the input portion contained the actual data from the dataset (e.g., CWE-ID, function before, function after, commit message etc.), and the output portion was the label of the data instance, indicating whether the commit mitigated or fixed a vulnerability. We fed this training data into the LLM so it could develop its reasoning capabilities for detecting vulnerabilities. Figure \ref{fig:finetuning}, gives an overview of the fine-tuning process.

The training configuration values of the model are given in the table \ref{tab:configuration}. The learning\_rate (2e-4) determines how much the model’s weights are updated with each step, while optimizer = adamw\_8bit reduces memory usage by using an 8-bit variant of AdamW. We set max\_steps to 30, after which training stops. The lora\_alpha value of 16 controls the scaling of LoRA adapter updates, and lora\_dropout (0.1) adds a small amount of dropout to prevent overfitting within LoRA. A weight\_decay of 0.01 provides regularization for the model’s parameters. The r value of 16 specifies the rank for LoRA’s low-rank matrices, and lr\_scheduler\_type = linear ensures the learning rate decreases linearly across the training process. The training setup, hyperparameters, prompts, data, and other relevant implementation details are provided for greater clarity and reproducibility.\footnote{\label{footnotee} \url{https://github.com/Hasan-Saju/llm-code-vulnerability}}.

\begin{table}[!htbp]
\caption{Fine-tuning Configuration Values}
\begin{center}
\begin{tabular}{|l|c|}
\hline
\textbf{Parameters} & \textbf{Values} \\
\hline
\textbf{learning\_rate} & 2e-4 \\
\hline
\textbf{optimizer} & adamw\_8bit \\
\hline
\textbf{max\_steps} & 30 \\
\hline
\textbf{lora\_alpha} & 16 \\
\hline
\textbf{lora\_dropout} & 0.1 \\
\hline
\textbf{weight\_decay} & 0.01 \\
\hline
\textbf{r} & 16 \\
\hline
\textbf{lr\_scheduler\_type} & linear \\
\hline
\end{tabular}
\label{tab:configuration}
\end{center}
\end{table}

\subsection{Dual-Agent Approach}

\begin{figure}[t]
\centerline{\includegraphics[width=0.49\textwidth]{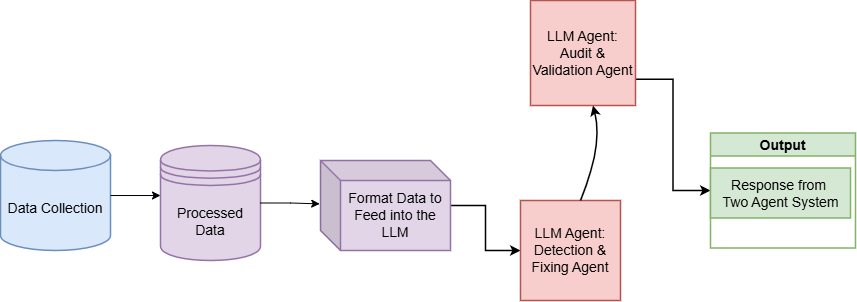}}
\caption{Dual-Agent LLM System Overview}
\label{fig:TwoAgent}
\end{figure}

We implemented a Dual-Agent system (Figure~\ref{fig:TwoAgent}) comprising two LLM agents with distinct roles: detection and validation. After collecting and pre-processing the data, the first agent is tasked with determining whether a given code change mitigates or fixes a vulnerability. It produces a response accompanied by its reasoning. This response is then passed to the second agent, which acts as an audit and validation agent.

The validation agent critically evaluates the output of the first agent, checking for logical inconsistencies, unsupported assumptions, or reasoning errors, and makes necessary revisions. This process helped mitigate common shortcomings of LLMs, including hallucinations and inconsistent output.

A key advantage of the Dual-Agent setup is its minimal resource requirement. Unlike SFT, which demands significant computational resources, and RAG, which necessitates constructing additional pipelines and external services like Chroma Database, the Dual-Agent approach offers a lightweight alternative. It improves reasoning quality and minimizes error propagation without requiring additional services or databases, as in RAG, or computational resources, as needed for fine-tuning.

\subsection{Environment}

All experiments were conducted on an NVIDIA T4 GPU (15 GB GPU memory), 32 GB of system RAM, and 112 GB of disk space. The base model used in our study is LLaMA 3.2-3B. For the SFT, RAG, and Dual-Agent approaches, we built upon this base model, applying the respective augmentations without altering its core architecture. This setup ensured consistent comparison across methods while keeping computational overhead manageable. 


\section{Results and Discussion}
\label{section:results}
We evaluated four techniques: Dual-Agent LLM, SFT, RAG and the Base Model across five CWEs using Accuracy, Precision, Recall, and F1 Score (Tables \ref{tab:rag_vertical}–\ref{tab:base_vertical}). 

\subsection{Findings}
RAG achieved the highest overall performance with an average F1 Score of 0.85, significantly outperforming SFT (0.80), Dual-Agent LLM (0.77), and the Base Model (0.67). Figure \ref{fig:F1Score} illustrates a comparison of the F1 scores of the different techniques across CWEs.

\begin{table}[!htbp]
\centering
\caption{Performance of RAG Across All CWEs}
\begin{tabular}{|l|c|c|c|c|}
\hline
\textbf{CWE} & \textbf{Accuracy} & \textbf{Precision} & \textbf{Recall} & \textbf{F1 Score} \\
\hline
CWE-119 & 0.80 & 0.73 & 0.89 & 0.80 \\
CWE-399 & 0.90 & 1.00 & 0.80 & 0.89 \\
CWE-264 & 0.85 & 0.80 & 0.89 & 0.84 \\
CWE-20  & 0.86 & 0.90 & 0.82 & 0.86 \\
CWE-200 & 0.90 & 1.00 & 0.80 & 0.89 \\
All (Avg)    & {0.86} & {0.88} & {0.84} & {0.85} \\
\hline
\end{tabular}
\label{tab:rag_vertical}
\end{table}

\begin{table}[!htbp]
\centering
\caption{Performance of SFT Across All CWEs}
\begin{tabular}{|l|c|c|c|c|}
\hline
\textbf{CWE} & \textbf{Accuracy} & \textbf{Precision} & \textbf{Recall} & \textbf{F1 Score} \\
\hline
CWE-119 & 0.80 & 0.89 & 0.72 & 0.80 \\
CWE-399 & 0.76 & 0.73 & 0.80 & 0.73 \\
CWE-264 & 0.85 & 0.89 & 0.80 & 0.84 \\
CWE-20  & 0.80 & 0.89 & 0.72 & 0.80 \\
CWE-200 & 0.84 & 0.89 & 0.80 & 0.84 \\
All (Avg)    & 0.81 & 0.85 & 0.76 & 0.80 \\
\hline
\end{tabular}
\label{tab:sft_vertical}
\end{table}

\begin{table}[!htbp]
\centering
\caption{Performance of Dual-Agent LLM Across All CWEs}
\begin{tabular}{|l|c|c|c|c|}
\hline
\textbf{CWE} & \textbf{Accuracy} & \textbf{Precision} & \textbf{Recall} & \textbf{F1 Score} \\
\hline
CWE-119 & 0.75 & 0.78 & 0.70 & 0.74 \\
CWE-399 & 0.85 & 0.82 & 0.90 & 0.86 \\
CWE-264 & 0.80 & 0.88 & 0.70 & 0.78 \\
CWE-20  & 0.75 & 0.78 & 0.70 & 0.73 \\
CWE-200 & 0.75 & 0.73 & 0.80 & 0.76 \\
All (Avg)     & 0.78 & 0.80 & 0.76 & 0.77 \\
\hline
\end{tabular}
\label{tab:dualagent_vertical}
\end{table}

\begin{table}[!htbp]
\centering
\caption{Performance of Base Model Across All CWEs}
\begin{tabular}{|l|c|c|c|c|}
\hline
\textbf{CWE} & \textbf{Accuracy} & \textbf{Precision} & \textbf{Recall} & \textbf{F1 Score} \\
\hline 
CWE-119 & 0.63 & 0.67 & 0.60 & 0.63 \\
CWE-399 & 0.73 & 0.72 & 0.80 & 0.76 \\
CWE-264 & 0.63 & 0.64 & 0.70 & 0.67 \\
CWE-20  & 0.68 & 0.75 & 0.60 & 0.67 \\
CWE-200 & 0.58 & 0.60 & 0.60 & 0.60 \\
All (Avg)   & 0.65 & 0.68 & 0.66 & 0.67 \\

\hline
\end{tabular}
\label{tab:base_vertical}
\end{table}

\begin{figure}[t]
\centering
\includegraphics[width=0.99\columnwidth]{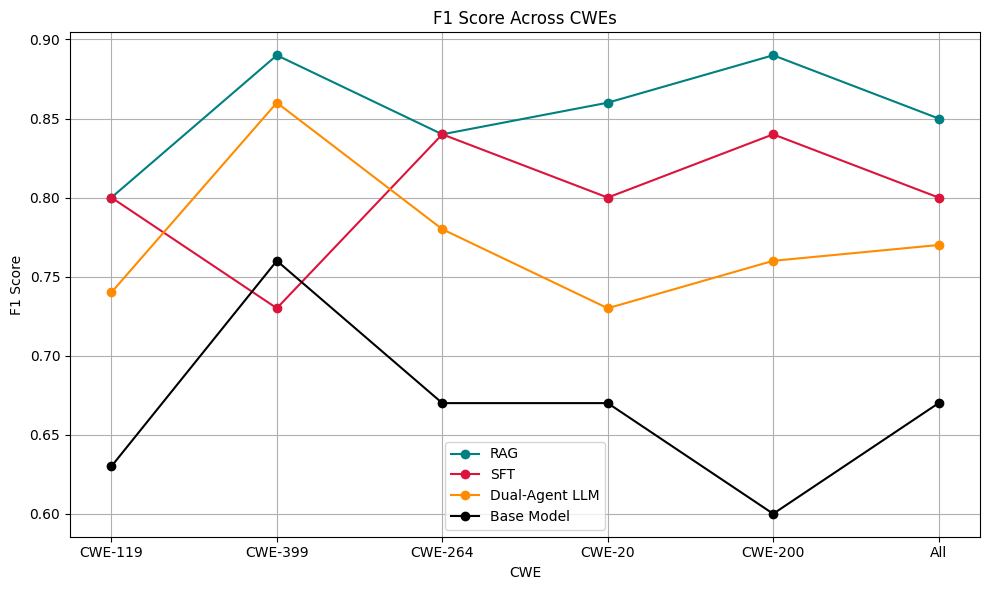}
\caption{F1-Score Comparison Among Techniques}
\label{fig:F1Score}
\end{figure}

For the Dual-Agent LLM, we observed that the first LLM agent often makes certain assumptions which affect its reasoning, sometimes resulting in incorrect responses, as shown in Figure \ref{fig:responseDual}. The second agent then helps identify any issues in the reasoning, as well as unwarranted assumptions or speculations, and works to correct them.

\begin
{figure}[t]
\centerline{\includegraphics[width=0.45\textwidth]{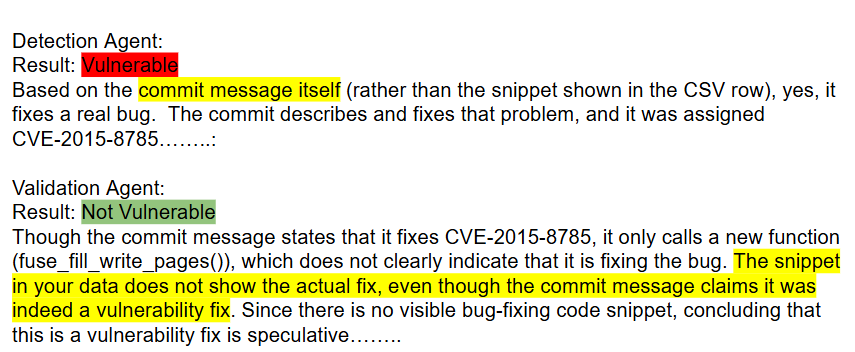}}
\caption{Response from Agent 1 and Agent 2 in the Dual-Agent LLM System}
\label{fig:responseDual}
\end{figure}

To assess whether the observed performance differences between models were statistically meaningful rather than due to random variation, we conducted paired t-tests on the F1 Scores across five CWE categories. This test is appropriate because it compares two related sets of measurements—each model's performance on the same CWEs—while accounting for within-category variance. By analyzing the mean difference in F1 Scores and calculating confidence intervals (CI), we can determine whether the performance gains of the advanced techniques over the Base Model are statistically significant.

A paired t-test on F1 Scores across CWE categories confirmed that all advanced techniques significantly outperformed the Base Model. \textbf{RAG} achieved the largest improvement ($t(4) = 5.97$, $p = 0.0042$, 95\% CI: [0.0975, 0.2292]), followed by \textbf{SFT} ($t(4) = 3.75$, $p = 0.0195$, 95\% CI: [0.0119, 0.1381]) and \textbf{Dual-Agent LLM} ($t(4) = 2.89$, $p = 0.0440$, 95\% CI: [0.0014, 0.1546]). These results indicate that all three techniques provide statistically significant improvements over the baseline model.

\subsection{Limitations}
While our results show promising improvements using RAG, SFT, and Dual-Agent approaches, certain limitations may affect the generalizability of the findings. Firstly, although our dataset is carefully curated from Big-Vul and real-world repositories before being balanced across five CWE categories, it could be further expanded to enhance model generalization and robustness. Despite its size, each instance was manually reviewed and selected from a widely used public dataset and enterprise-grade codebases, ensuring high label fidelity and practical relevance. Secondly, all models were evaluated on a static set of five CWE types. This choice reflects the higher availability of these CWE categories in enterprise codebases compared to others, but it may limit generalization to less frequently occurring vulnerability types. Thirdly, our proprietary dataset includes code from a single research partner, which may introduce source bias. Future work will aim to scale the dataset, include more CWEs, validate across broader codebases and evaluate the performance in broader settings.

\section{Conclusion} \label{section:conclusions}
In this paper, we conducted a comprehensive exploration and empirical evaluation of three LLM-based methods, Retrieval-Augmented Generation (RAG), Supervised Fine-Tuning (SFT), and a Dual-Agent LLM system, alongside a baseline LLM model. Our target was to compare their respective capabilities to detect software code vulnerabilities. We generated a dataset from multiple sources, covering five key CWEs (CWE-119, CWE-399, CWE-264, CWE-20, and CWE-200). We found that RAG achieved the highest overall performance with an accuracy of 0.86 and an F1-score of 0.85. Through this study, we aimed to provide a holistic comparison of different LLM-based vulnerability detection approaches to aid researchers or developers with choosing the right technique for their applications. To the best of our knowledge, this paper involved the first successful attempt to apply a Dual-Agent LLM in the domain of code vulnerability detection. Not only that, but the dataset created for this study addresses the issue of class imbalance gap faced by related work in this domain. In the future, we plan to expand the dataset with more CWE types and additional instances per category. Our findings show that RAG helps reduce reasoning errors through domain knowledge, SFT improves detection via targeted learning, and the Dual-Agent setup enhances detection capability by validating the response of the first agent. Overall, these methods significantly improve vulnerability detection over the base model.


\vspace{12pt}
\color{red}

\end{document}